\documentclass[useAMS,usenatbib,usegraphicx]{mn2e}

\usepackage{graphicx}
\DeclareGraphicsExtensions{.pdf,.png,.jpg,.ps}
\usepackage{epsfig}

\usepackage[bottom]{footmisc}
\usepackage{color,amsmath,enumitem}%,breqn}
\usepackage{threeparttable}
\usepackage{graphicx}
\DeclareGraphicsExtensions{.pdf,.eps,.jpg,.ps,.png}
\usepackage{epsfig}
\usepackage{url}
\usepackage{ amssymb }
\usepackage{pdflscape}
\usepackage{afterpage}
\usepackage{capt-of}
\usepackage{gensymb}

\title[OGLE RR Lyraes in the Magellanic Bridge]{The Properties of the Magellanic Bridge Based on OGLE IV RR Lyrae Variables}
\author[R. Wagner-Kaiser, A. Sarajedini]
  {R.~Wagner-Kaiser,$^1$\thanks{Email: rawagnerkaiser@astro.ufl.edu}
  Ata~Sarajedini$^1$\thanks{Email: ata@astro.ufl.edu}
\\
  $^1$University of Florida, Department of Astronomy, 211 Bryant Space Science Center, Gainesville, FL, 32611 USA\\
}

\begin{document}

\date{}

\pagerange{\pageref{firstpage}--\pageref{lastpage}} \pubyear{2002}

\maketitle

\label{firstpage}

%% Abstract and Keywords

\begin{abstract}
We examine the properties of the Magellanic Bridge connecting the Large and Small Magellanic Clouds using ab-type RR Lyrae variables from the extensive dataset of the Optical Gravitational Lensing Experiment (OGLE), Phase IV data release. The metallicities of the RR Lyraes are determined from the characteristics of their light curves, with an average abundance of [Fe/H] = -1.790 $\pm$ 0.011 (sem) in the Magellanic Bridge.  From the individual reddenings of these stars, derived via their minimum light curve colors, we determined a median color excess of E(V--I) = 0.101 $\pm$ 0.007 (sem) (implying E(B-V) $\approx$ 0.077). The peak distance modulus of 18.57 $\pm$ 0.048 (sem) places the Bridge stars at distances between the two systems. The metallicity and distance distributions probe the structure of the Magellanic system as a whole, revealing a smooth transition that connects the galaxies. An examination of the HI content does not find a clear correlation between HI emission strength and RR Lyrae spatial distribution, suggesting that the old stellar populations may trace the overlapping halo distributions of the two Magellanic Clouds.
\end{abstract}


%% Introduction

\section{Introduction}\label{intro}
The Magellanic clouds are an important piece of the puzzle to help us better understand galaxy interactions, as well as the dynamical and star formation history of our own Galaxy. Not only do the gaseous features surrounding the Large and Small Megallanic Clouds (L/SMC) suggest past interactions between the two galaxies, but they also indicate the presence of a common HI envelope. However, there are still disagreements about the nature of these interactions, and the role of the Magellanic Bridge (MB) connecting the two galaxies. Some studies suggest that interactions between the LMC and SMC stem from multiple pericentric passages of the galaxies as they orbit our Galaxy, with the inter-galactic features resulting from tidal stripping and ram pressure (\citealt{Ruuzicka:2009}). Other work suggests that the Magellanic clouds have only begun interacting with each other and the Milky Way during two recent encounters, estimated to have occurred $\sim$2 Gyr and $\sim$250 Myr ago (\citealt{Kallivayalil:2013, Noel:2015}).

Proper motion and internal kinematics suggest that the MB likely resulted from a recent direct collision of the two galaxies (\citealt{Kallivayalil:2013, Besla:2012, Diaz:2012}). Such an interaction is thought to have triggered star formation in the MB, as pointed out by \cite{Grondin:1992, Mizuno:2006, Skowron:2014, Chen:2014}. However, based on their ages, the young stars could have either formed \textit{in situ} in the MB, or formed in the SMC and been tidally stripped from it in the past $\sim$ 200 Myr (\citealt{Noel:2015}).

\cite{Skowron:2014} find that although the young stellar populations in the MB tend to be concentrated near the SMC, there is evidence for a continuity of a young population throughout the Bridge region, consistent with the HI densities. Intermediate age stellar populations have also been identified in the Bridge region, although they appear to occupy a distinctly different physical distribution than the pockets of star formation activity (\citealt{Noel:2013, Skowron:2014, Noel:2015}). In particular, \cite{Skowron:2014} find a non-uniform distribution of red giant and asymptotic giant branch stars across the region between the LMC and SMC.

Work by \cite{Borissova:2004,Borissova:2006} suggests the existence of a stellar halo in the LMC based on velocity dispersion measurements of RR Lyrae variables, located as much as 2.5 degrees away from the center of the LMC. The large velocity dispersion of these stars ($\sigma_{RV}$ = 50 $\pm$ 2 km s$^{-1}$) indicates a 
homogeneous, old, and metal-poor halo. Other recent work has offered star counts as evidence that the LMC disk may extend to $\sim$16 kpc (\citealt{Saha:2010, Besla:2016}).

We now turn our attention to the tools we plan to employ in the present study. RR Lyrae variables provide valuable insights into the early formation epochs of galaxies (\citealt{Sarajedini:2011,Smith:1995}). Pulsating on the horizontal branch, these variable stars are quite old ($\gtrsim$ 10 Gyr), allowing us to look more closely at the ancient population of the Magellanic Bridge (\citealt{Alcock:2000,Soszynski:2009}). From the light curve shapes and characteristics of these stars, it is possible to determine not only distances, but also chemical properties and line-of-sight extinctions. The confluence of these many fundamental properties allows us a deep look into the history and evolution of their host galaxies.

The Optical Gravitational Lensing Experiment (OGLE, \citealt{Udalski:2012}) has dramatically increased the number of RR Lyrae stars (and other types of variables) identified in the Magellanic system. Recent work has extended OGLE to even greater coverage of the LMC and SMC in Phase IV of the survey and now catalogs over 45,000 confirmed RR Lyrae variables (\citealt{Soszynski:2016}), now including fields in the MB. While many previous studies have examined the RR Lyrae populations from the OGLE survey (\citealt{Pejcha:2009, Kinman:1991, Haschke:2011, Haschke:2012a, Haschke:2012, Wagner-Kaiser:2013, Deb:2014, Inno:2016}, among others), the recent expansion of spatial coverage allows a more detailed look at the old populations of the MB.

Here, we present a study of the reddening, metallicity, distance, and spatial distribution of RR Lyrae ab-type stars in the Magellanic Bridge between the LMC and SMC galaxies. Section \ref{Data} describes the dataset of the OGLE database. An analysis of the metallicities, distances, and reddenings of the RR Lyrae stars is presented in Section \ref{Results}. Section \ref{bridge} examines the HI content in the MB compared to the distribution of reddening and RR Lyrae positions. We conclude in Section \ref{conclusions}.

%%%%%%%%%%%%%%%%%%%%% Data/Methods

\section{Data}\label{Data}

In 1992, OGLE began with Phase I, with additional coverage (Phases II and III) released in following years (\citealt{Udalski:2012, Udalski:2003, Soszynski:2009}). The most recent dataset, OGLE Phase IV, extended the temporal and spatial coverage using the 1.3 meter Warsaw telescope at Las Campanas Observatory between March 2010 and July 2015 (\citealt{Soszynski:2016}). Almost 500 fields of size 35 by 35 arcmin cover $\sim$650 square degrees of the sky in Cousins I-band and Johnson V-band filters, with the majority of the observations in the I-band.

Full details of the reduction procedures, photometric calibrations, and astrometric transformations may be found in \cite{Udalski:2008} and \cite{Udalski:2015}. Specifically, difference imaging techniques were employed to detect probable variables with periods between 0.2 and 1 day. Fourier decomposition and template fitting to the I-band light curves were used to preselect candidates, and periods were determined with a Fourier analysis (\citealt{Soszynski:2016}).

\begin{figure}
  \centering
    \includegraphics[width=0.5\textwidth]{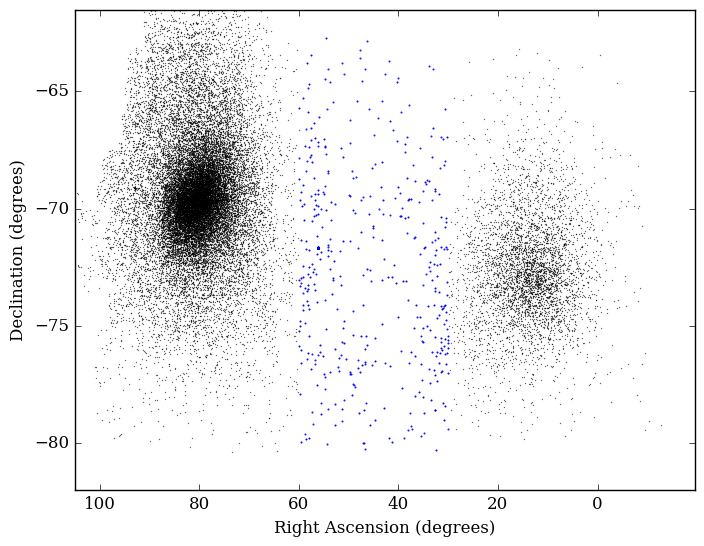}
  \caption{The RR Lyrae in OGLE classified as ab type are shown as black points, with the RR Lyrae falling into the Bridge region, indicated as blue points.}
  \label{OGLEcov}
\end{figure}

The sample of ab-type RR Lyrae variables consists of 26,167 putative LMC stars and 4,750 in the SMC. The samples overlap in the area of the MB, as shown in Figure \ref{OGLEcov}. In this figure, we have identified what we consider to be the MB variables as the blue points with the distribution of all identified OGLE RR Lyrae ab-type stars shown for comparison. We determine the Bridge region by eye to fall between the right ascensions of 30$^{\circ}$ and 60$^{\circ}$ (in the range of 2$^{h}$ and 4$^{h}$). There are 467 ab-type RR Lyrae stars in this region with well-determined magnitudes, I-band amplitudes, and periods; 452 of these stars also have well-determined Fourier parameters.

%%%%%%%%%%%%%%%%%%%%% Results section

\section{Results}\label{Results}

In Section \ref{metallicities}, OGLE data are used to compare metallicities of ab-type RR Lyrae aross the MB region. We derive extinctions for the RR Lyrae in Section \ref{extinction}, determined from the minimum light color of their light curves. The individual stellar reddenings are leveraged to examine the distribution of RR Lyrae distances in Section \ref{distances}.

%%%%%%%%%%%%%%%%%%%%% Results: Metallicities

\subsection{Metallicities}\label{metallicities}

\subsubsection{RRab Stars}\label{metalsRRab}

Metallicities for the RRab stars in our sample are found with the relationship from \cite{Alcock:2000} shown in equation \ref{eq1}, which is on the \cite{Zinn:1984} abundance scale. P$_{ab}$ represents the period of the ab-type variables and V$_{amp}$ is the light curve amplitude in the V-band:

\begin{equation} \label{eq1}
[Fe/H]_{ab} = -8.85 [log(P_{ab})+0.15 V_{amp}] - 2.6.
\end{equation}

\noindent However, to use equation \ref{eq1}, it is necessary to convert the I-band amplitudes from the OGLE IV catalog to V-band amplitudes. As in \cite{Wagner-Kaiser:2013}, we make use of equation 1 in \cite{Dorfi:1999}:

\begin{equation} \label{eqn2}
V_{amp} = 0.075 + 1.497I_{amp},
\end{equation}

\noindent where V$_{amp}$ and I$_{amp}$ are the light curve amplitudes  in the V- and I-band, respectively. \cite{Dorfi:1999} derive this empirical relation with a R$^2$ correlation coefficient of 0.904. 

While \cite{Alcock:2000} estimate an error of $\sim$0.31 dex per star from Equation \ref{eq1}, the actual error may be less (\citealt{Jeffery:2011}). Blazhko effects, which cause changes in amplitude over time, may affect some portion of RR Lyrae stars and incorporate errors up to 0.03 dex (\citealt{Kunder:2010}). Additional errors of up to 0.014 dex may be introduced by the conversion of the I-band amplitude to the V-band value (\citealt{Wagner-Kaiser:2013}). However, these two additional sources of error are negligible when compared to the overall error from \cite{Alcock:2000}.

For comparison, the metallicity is also derived from Fourier light curve decomposition parameters from the OGLE catalog via the I-band photometry. The methodology and derivation of the metallicity via Fourier parameters from \cite{Smolec:2005} has been recently updated by \cite{Skowron:2016}. We use this newer calibration of [Fe/H] to calculate the metallicity on the \cite{Jurcsik:1995} metallicity scale:

\begin{equation} \label{eq3}
[Fe/H]_{J95, ab} =  2.132 - 5.394 P_{ab} - 1.009 \phi_{31} + 0.164 \phi_{31}^{2} 
\end{equation} 

\noindent where P$_{ab}$ is the period of the RRab variables and $\phi_{31}$ is a Fourier parameter. A phase change must be accounted for in $\phi_{31}$; the OGLE Fourier decomposition uses a cosine series, while \cite{Skowron:2016} uses a sine Fourier series. [Fe/H] is converted from Equation \ref{eq3} from the \cite{Jurcsik:1995} metallicity scale to the \cite{Zinn:1984} scale (ZW):

\begin{equation} \label{eq4}
[Fe/H]_{ZW, ab} = 1.028 [Fe/H]_{J95, ab} - 0.242,
\end{equation}

\noindent from \cite{Papadakis:2000}.

\begin{figure*}
  \centering
    \includegraphics[width=\textwidth]{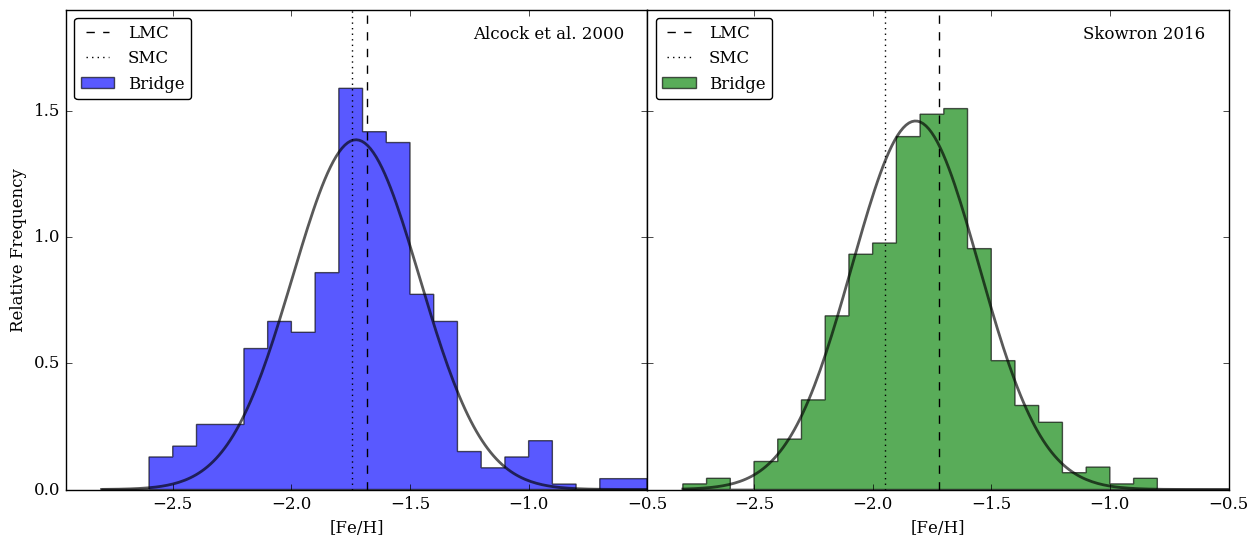}
  \caption{The distribution of RR Lyrae metallicities in the Magellanic Bridge determined from the \protect\cite{Alcock:2000} relation (A00, left) and those from \protect\cite{Skowron:2016} (S16, right). In each panel, Gaussians (black solid lines) are fit to each of the distributions. The peak LMC and SMC metallicities from these two approaches are indicated in each panel as dashed and dotted vertical lines, respectively. In both panels, the histograms are normalized to an area of 1.}
  \label{metallicityhist}
\end{figure*}

In Figure \ref{metallicityhist}, the results are presented from our determination of [Fe/H] from Equations \ref{eq1} through \ref{eq4} for ab-type variables in the MB. In the left panel, the metallicity distribution from the \cite{Alcock:2000} relation is presented (A00, blue), while the results from the \cite{Skowron:2016} equation are shown in the right panel (S16, green). Gaussian profiles are fit to each of the metallicity distributions in Figure \ref{metallicityhist}. The derived peak metallicities and errors representing the standard error of the mean are given in Table \ref{tab:metallicities}. The peak metallicity of the Bridge stars falls between the mean LMC and SMC metallicities.

The metallicities derived via the Fourier parameter method from \cite{Skowron:2016} tend towards more metal-poor values, although the 3-$\sigma$ errors from the two results marginally overlap. For comparison, metallicities from the literature are presented in Table \ref{tab:metallicities_lit}. Also included in Figure \ref{metallicityhist} for comparison are vertical lines indicating the peaks of the LMC (dashed) and SMC (dotted) metallicity distributions from the OGLE dataset.

\begin{table}
\centering
    \begin{minipage}{180mm}
    \caption{Peak Metallicity of Magellanic Bridge RR ab Lyrae}
    \begin{tabular}{@{}ccc@{}}
    \hline
\textbf{[Fe/H]}  & \textbf{Method} & \textbf{Reference}  \\  
\hline
-1.744 $\pm$ 0.016		& Equations \ref{eq1} \& \ref{eqn2}	& \cite{Alcock:2000}  \\
-1.835 $\pm$ 0.014		& Equations \ref{eq3} \& \ref{eq4}	& \cite{Skowron:2016} \\
        \hline
    \label{tab:metallicities}
    \end{tabular}
    \end{minipage}
\end{table}

\begin{table*}
\centering
    \begin{minipage}{180mm}
    \caption{RR Lyrae Literature Metallicities}
    \begin{tabular}{@{}cccc@{}}
    \hline
\textbf{Galaxy} & \textbf{[Fe/H]} & \textbf{Method} &  \textbf{Reference}  \\  
\hline
LMC	& -1.70 $\pm$ 0.25	 & RRab, \cite{Alcock:2000}		& \cite{Wagner-Kaiser:2013} \\
LMC	& -1.62 $\pm$ 0.10	 & RRab, \cite{Smolec:2005}	& \cite{Wagner-Kaiser:2013} \\
LMC	& -1.50 $\pm$ 0.24	 & RRab, \cite{Smolec:2005}	& \cite{Haschke:2012a} \\
LMC	& -1.63 $\pm$ 0.095	& Spectra					& \cite{Borissova:2006} \\ \hline
SMC	& -1.89 $\pm$ 0.03	& RRab, \cite{Smolec:2005}		& \cite{Kapakos:2012} \\
SMC & -1.70 $\pm$ 0.27	& RRab, \cite{Smolec:2005}		& \cite{Haschke:2012a} \\
        \hline
    \label{tab:metallicities_lit}
    \end{tabular}
    \end{minipage}
\end{table*}

%%%%%%%%%%%%%%%%%%%%% Results: Metallicity Gradient

\subsubsection{Metallicity Structure}\label{metallicitygrad}

From the abundances derived in Section \ref{metallicities}, the metallicity of the RR Lyrae stars can be examined in the Bridge and across the Magellanic system. For each ab RR Lyrae variable in the Magellanic system, we calculate its projected radius: 

\begin{eqnarray} \label{angdist}
\cos{\theta}=[\sin{(90-\delta)}*\sin{(90-\delta_{galx})}*\cos{(\alpha-\alpha_{galx})}]  \\  \nonumber
+[\cos{(90-\delta)}*\cos{(90-\delta_{galx})}] ,  \\ 
\end{eqnarray}

\noindent from \cite{vanderMarel:2001}, where  $\alpha_{galx}$ and $\delta_{galx}$ are the right ascensions and declinations of the LMC center (5$^h$ 23$^m$ 34.5$^s$, -69$^{\circ}$ 45$^{`}$ 22$^{``}$,\citealt{vanderMarel:2001}) and SMC center (0$^h$ 55$^m$ 7.2$^s$, -72$^{\circ}$ 47$^{'}$ 59$^{''}$, \citealt{Gonidakis:2009}). The variables $\alpha$ and $\delta$ are the right ascension and declination of the individual OGLE RR Lyrae stars. This is converted to a physical distance assuming an LMC distance modulus of 18.49 (\citealt{de-Grijs:2014}) and an SMC distance modulus of 18.96 (\citealt{de-Grijs:2015}).

\begin{figure}
  \centering
    \includegraphics[width=0.48\textwidth]{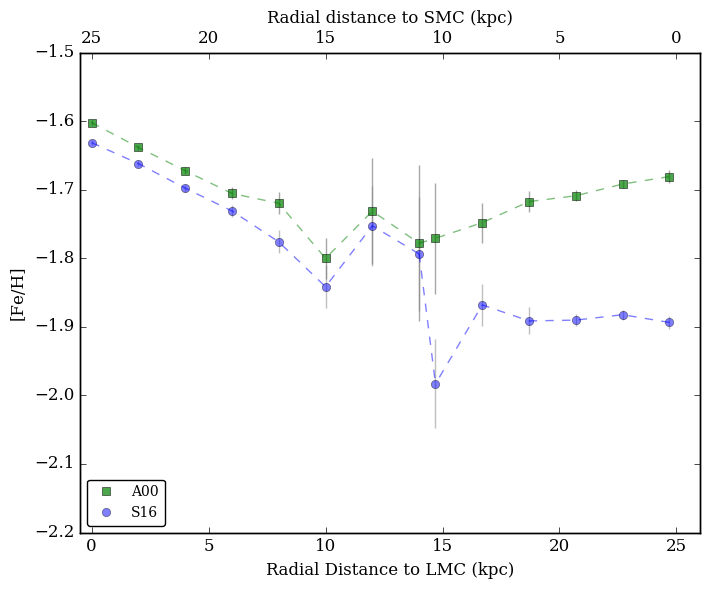}
  \caption{The metallicity binned every 2 kpc in distance from the center of the LMC on the lower abscissa and the center of the SMC from the upper abscissa. Metallicities are plotted for the \protect\cite{Alcock:2000} (green) and \protect\cite{Skowron:2016} (blue). The errorbars represent the standard error of the mean in each bin.}
  \label{feh_structure}
\end{figure}

Figure \ref{feh_structure} traces the metallicity changes across the Magellanic system. The stars are binned at intervals of 2 kpc and the mean and standard errors of the metallicities of the RR Lyrae are determined in each bin. After reaching an apparent minimum in the region of the Bridge, the overall metallicity becomes marginally more metal-rich near the SMC for the \cite{Alcock:2000} metallicities and flattens out for the \cite{Skowron:2016} derivation. The overall structure of the system is examined further in Sections \ref{distances} and \ref{bridge}.

%%%%%%%%%%%%%%%%%%%%% Results: Extinction Mapping

\subsection{Extinction}\label{extinction}

\cite{Guldenschuh:2005} found the minimum light color of ab-type RR Lyrae variables, the color at the faintest point of the light curve, to be equal to $(V-I)_o$ = 0.58 $\pm$ 0.02 regardless of their intrinsic properties. This relationship is used to determine individual reddenings for the RR ab stars in our LMC and SMC samples. From \cite{Guldenschuh:2005}, the E(V-I) reddening of an RR Lyrae ab star can by calculated by:

\begin{equation} \label{eq6}
E(V-I) = (V-I)_{min} - 0.58.
\end{equation}

\noindent If it is assumed that the ab-type RR Lyraes light curves are sinusoidal and symmetric, we can calculate the minimum light color for each star as:

\begin{equation} \label{eqn7}
(V-I)_{min} = [<V> + \frac{A_V}{2}] - [<I> + \frac{A_I}{2}],
\end{equation}

\noindent where A$_{V}$ and A$_{I}$ are the V- and I-band amplitudes of the light curves.

However, because the shapes of ab-type RR Lyrae light curves are asymmetric and non-sinusoidal, this relation must be modified somewhat. From simulations, \cite{Wagner-Kaiser:2013} determined a correction to be applied to the minimum light color in Equation \ref{eqn7} using RRab light curve templates from \cite{Layden:2000} as:

\begin{equation} \label{eqn8}
\Delta = (V-I)_{min}^{template} - [<V> + \frac{A_V}{2}] - [<I> + \frac{A_I}{2}].
\end{equation}

\noindent finding a mean value of $\Delta$ for the 6 ab-type templates to be $\langle$$\Delta$$\rangle$ = -0.061 $\pm$ 0.017. We apply this correction to the minimum light colors derived from Equation \ref{eqn7} to account for the typically sawtooth, asymmetric ab-type light curves.

\begin{figure}
  \centering
    \includegraphics[width=0.48\textwidth]{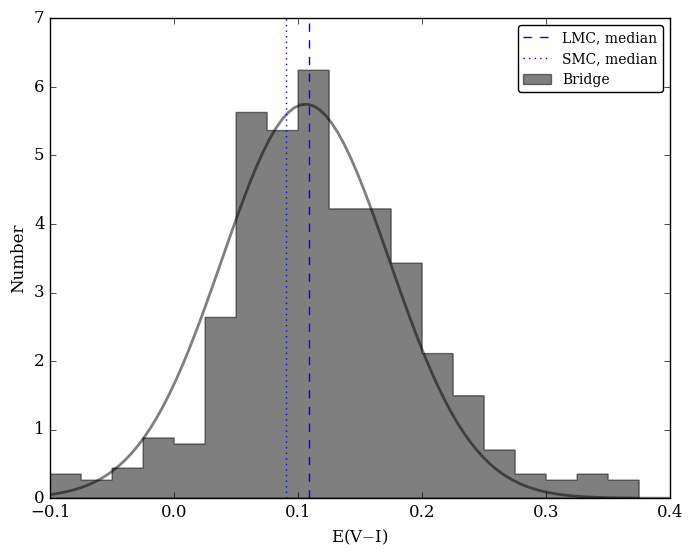}
  \caption{Histogram of  reddening values for ab-type RR Lyraes in the Bridge, normalized to an area of 1. The solid line is the Gaussian fit to the central portion of the distribution. The peaks of the reddening distributions for the LMC and SMC are shown as dashed and dotted lines, respectively.}
  \label{redhist}
\end{figure}

In Figure \ref{redhist}, the reddening distribution for the RRab stars in the MB is shown. A small percentage of the RR ab-type variables ($\sim$5\%) are found to have negative reddening values - a gratifyingly minimal amount. A Gaussian profile is fit to the reddening distribution, with a peak at E(V--I) of 0.101 $\pm$ 0.007, equivalent to E(B-V) of 0.077 assuming R$_V$ = 3.1. The error in each individual reddening is $\sigma$=0.026 mag from the uncertainty in $(V-I)_{o,min}$ and the non-sinusoidal correction. Table \ref{tab:red_lit} shows a comparison of our reddening results for the LMC and SMC with those in the literature.

While the approach of \cite{Guldenschuh:2005} rests on the somewhat contested presumption of an absolute minimum light color for RR Lyrae (supporting views: \citealt{Kunder:2010, Bhardwaj:2014}, opposing: \citealt{Collinge:2006, Layden:2013}), our estimates of the peak reddening from this approach fit well with other methods of reddening determination (see Table \ref{tab:red_lit}).

Upon examination of Figure \ref{redhist}, the reddening appears to be largely a continuous distribution. There is an overabundance beyond the Gaussian fit of the Bridge RR Lyrae for E(V--I) $\gtrsim$ 0.15, which could be indicative of internal reddening affecting some Bridge stars. However, due to the uncertainties associated with reddening calculations and photometric uncertainties for individual stars ($\approx$0.026), it is difficult to disentangle. There is no clear bi-modality to separate the RR Lyrae affected by foreground reddening only and those affected by both foreground and internal reddening. We return to this in Section \ref{bridge}.

\begin{table*}
\centering
    \begin{minipage}{180mm}
    \caption{Literature Reddening}
    \begin{tabular}{@{}cccc@{}}
    \hline
Galaxy & E(V--I) & Method &  Reference  \\  
\hline
LMC  &	0.12 $\pm$ 0.05	 & 	RRab OGLE III		& \cite{Wagner-Kaiser:2013} \\
LMC  &	0.11 $\pm$ 0.06 & 	RRab OGLE III		& \cite{Haschke:2011}\\
LMC  &	0.09 $\pm$ 0.07	 & 	Red Clump			& \cite{Haschke:2011} \\
LMC  &	0.11 $\pm$ 0.06	& 	LMC Clusters		& \cite{Wagner-Kaiser:2013} (average from \\
	&				&					& \cite{Johnson:1999, Olsen:1998, Brocato:1996})  \\ \hline
SMC  &	0.07 $\pm$ 0.06 & 	RRab OGLE III		& \cite{Haschke:2011}\\
SMC  &	0.04 $\pm$ 0.06 & 	Red Clump			& \cite{Haschke:2011} \\
        \hline
    \label{tab:red_lit}
    \end{tabular}
    \end{minipage}
\end{table*}

%%%%%%%%%%%%%%%%%%%%% Results: Distances

\subsection{Distances}\label{distances}

\subsubsection{RRab Stars}\label{distsRRab}

To determine individual distances to the RR ab variables, the relation from \cite{Chaboyer:1999} is used to relate the absolute magnitude to the metallicity of the stars: $M_V = (0.23\pm0.04)([Fe/H] + 1.6) + (0.56\pm0.12)$. We calculate the absolute magnitudes of the RR Lyrae with the OGLE mean V-band magnitudes, using the individual extinctions of each star (see Section \ref{extinction}).

\begin{figure*}
  \centering
    \includegraphics[width=\textwidth]{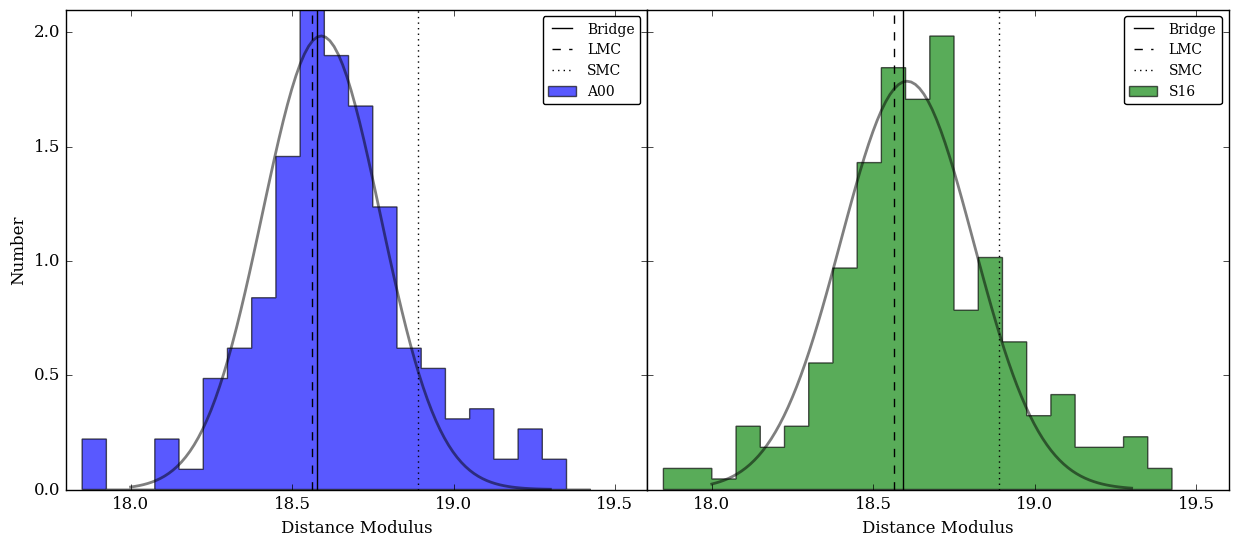}
  \caption{Distances derived from the \protect\cite{Alcock:2000} and \protect\cite{Skowron:2016} metallicity determinations are in blue and green, respectively, for the RR ab stars. The solid curve is the Gaussian distribution fitted to the distances of the Bridge RR Lyrae, with the peak indicated by the solid vertical line. The resulting Gaussian peaks are provided in Table \ref{tab:dists} with the corresponding errors. In both panels, the histograms are normalized to an area of 1, and vertical lines indicating the median LMC (dashed) and SMC (dotted) distances from the OGLE RR Lyrae are plot for comparison.}
  \label{disthist}
\end{figure*}

The \cite{Alcock:2000} and \cite{Skowron:2016} derivations of the metallicity are both used to derive distances and the resulting distributions for the stars in the MB are compared in Figure \ref{disthist}. Gaussian profiles are fit to each distribution, with peaks for each provided in Table \ref{tab:dists}. The results from the two metallicity determinations are consistent exhibiting a difference of only 0.011 mag; the distances agree within the 1-$\sigma$ errors.

The peak distance of the RR Lyrae in the Bridge region is much more similar to the LMC than the SMC; this may suggest the Bridge RR Lyrae are dominated by stars more likely to be associated to the LMC. We see less structure in the distance distribution derived via the \cite{Alcock:2000} relation compared to \cite{Skowron:2016}. The latter has a possible secondary peak around the distance of the SMC at $\mu$ $\approx$ 18.9. The primary peak at the approximate distance of the LMC may again be indicative that the stars in the Bridge are largely from the LMC. However, a secondary peak is not clear in the Bridge distances derived via the \cite{Alcock:2000} relation, though there tends to be an overabundance at higher distances.

\begin{table}
\centering
    \begin{minipage}{180mm}
    \caption{Peak Distance of Magellanic Bridge RR ab Lyrae}
    \begin{tabular}{@{}cc@{}}
    \hline
%\textbf{Galaxy} & \textbf{RRL type} & 
\textbf{Distance Modulus} & \textbf{Method}  \\  
\hline
18.591 $\pm$ 0.067	&  \cite{Alcock:2000}  \\
18.603 $\pm$ 0.068	& \cite{Skowron:2016} \\
        \hline
    \label{tab:dists}
    \end{tabular}
    \end{minipage}
\end{table}

\subsubsection{Structure of the Magellanic Bridge}\label{diststruct}

\begin{figure}
  \centering
    \includegraphics[width=0.48\textwidth]{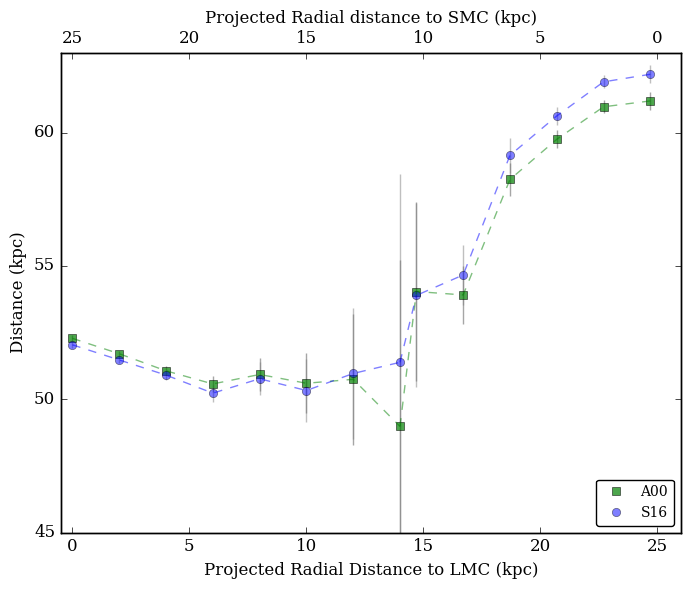}
  \caption{The mean distance of the RR ab variables along their projected radial distance in kpc from the center of the LMC-SMC system. The LMC is shown along the lower abscissa and the SMC is along the upper abscissa. We plot the distances derived from the \protect\cite{Alcock:2000} metallicity derivation in green and those from \protect\cite{Skowron:2016} in blue. For both derivations, there is a smooth transition of stellar distances that appears to connect the two galaxies in the Magellanic Bridge area. The errorbars represent the standard error of the mean in each bin.}
  \label{dist_structure}
\end{figure}

The derived distances from Section \ref{distances} are used to examine the line-of-sight distribution of the ab-type RR Lyrae stars using their projected radial distances, binned every 2 kpc. As seen in Figure \ref{dist_structure}, the result shows a clear and smooth transition of stars between the two galaxies. This could be indicative of the overlapping halo distributions belonging to the LMC and SMC, as discussed further in Section \ref{bridge}.

%%%%%%%%%%%%%%%%%%%% Conclusion

\section{The Magellanic Bridge}\label{bridge}

\subsection{Extinction Analysis}

As noted in Section \ref{intro}, the Magellanic Clouds are a key component in understanding the early formation and evolution of the Milky Way system. The LMC and SMC are considered typical dwarf galaxies similar to the type that may have been accreted to form more massive galaxies, such as our own. Unlocking the history of the Magellanic Clouds thus helps shed light on the possible growth mechanisms of larger spiral galaxies as well as the local history of the Milky Way Galaxy and its satellites. By examining the Bridge between the LMC and SMC, we gain insight into the structure of the system and their history of interactions.

Studies have shown that the young and intermediate age stars in the MB have a clumpy, non-uniform distribution (\citealt{Noel:2013,Noel:2015,Skowron:2014}). The asymmetric locations of these stellar populations suggest that either they formed in situ or were stripped from the LMC or the SMC during an interaction. However, the older RR Lyrae don't show any clear structure in their locations across that region (e.g.: Figure \ref{OGLEcov}). If the older populations, such as RR Lyrae, follow the trends of their younger counterparts, it would give credence to tidal stripping (as in-situ formation is not likely at such an early time in the Bridge). However, if the distribution of RR Lyrae is smooth across the Bridge, it may be indicative of an extended halo distribution of the SMC and LMC overlapping in the Bridge (\citealt{Besla:2016,Saha:2010}).

\begin{figure}
  \centering
    \includegraphics[width=0.5\textwidth]{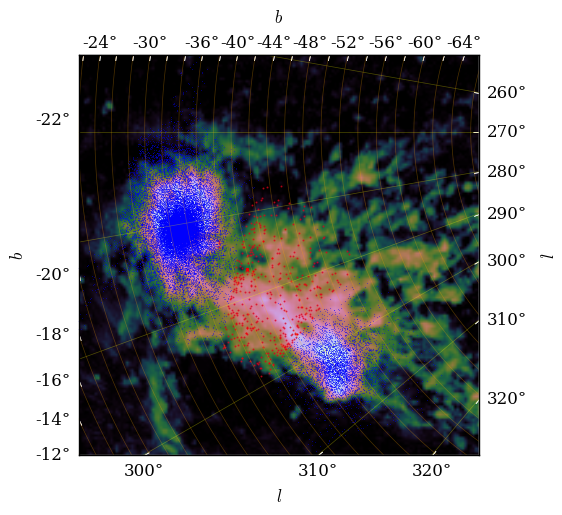}
  \caption{HI integrated intensity from \protect\cite{Putman:2003}, overlaid with RR Lyrae locations from the OGLE survey indicated in blue. The RR Lyrae that fall into the Bridge region are indicated by the red points.}
  \label{RRLHI}
\end{figure}

In Figure \ref{RRLHI}, we use HI observations of the LMC, SMC, and MB to examine RR Lyrae positions with respect to the HI integrated intensity (\citealt{Putman:2003}). Visual inspection suggests there is not strong association between the physical distribution of the RR Lyrae and greater HI intensity.

To delve in further, the HI content of the region from \cite{Putman:2000} is compared to the extinction traced by the RR Lyrae (as per Section \ref{extinction}). The integrated HI intensity from Figure \ref{RRLHI} is converted to a E(V--I) reddening through the following process. From \cite{Fukui:2014}, we can use Equation \ref{eq10} to convert the integrated HI intensity, I$_{HI}$, to the column density, N$_{HI}$:

\begin{equation} \label{eq10}
N_{HI} (cm^{-2}) = (1.823 \times 10^{18}) \times I_{HI} (K km s^{-1})
\end{equation}

\noindent Following this, the E(B--V) is calculated from the HI column density using the relation from \cite{Bohlin:1978}and \cite{Diplas:1994}:

\begin{equation} \label{eq11}
N_{HI} (cm^{-2})= (4.93 \pm 0.09) \times 10^{18} E(B-V)
\end{equation}

\noindent Which is converted to E(V--I) with the relation 1.32$\times$E(B--V), giving us an estimate of the amount of reddening caused by HI emission.

\begin{figure}
  \centering
    \includegraphics[width=0.5\textwidth]{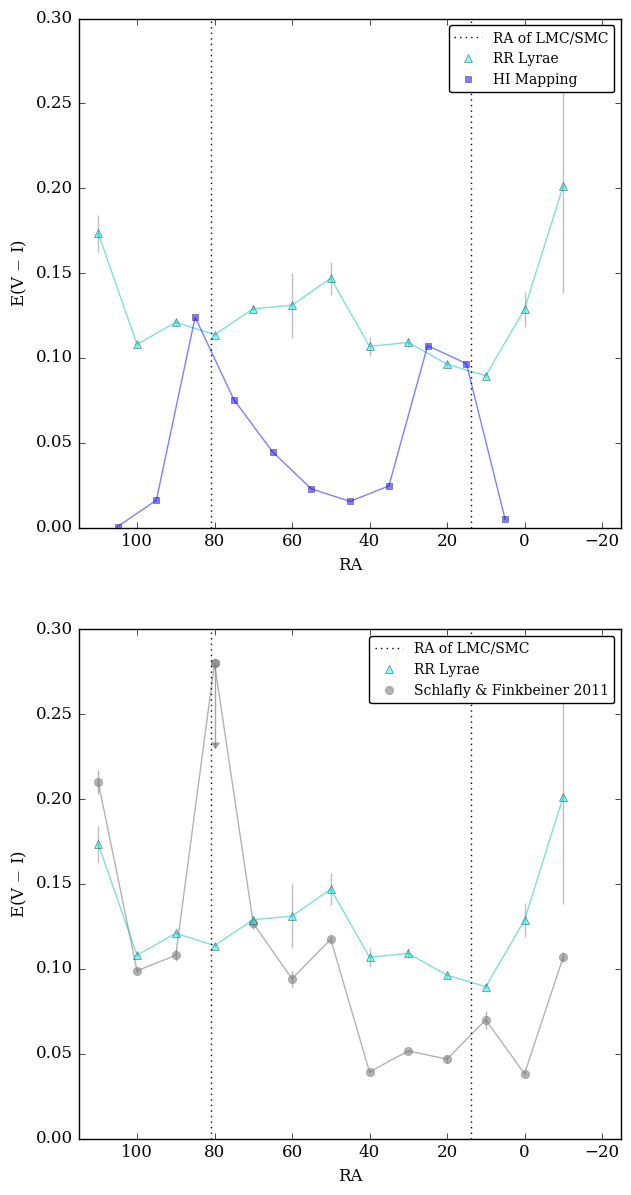}
  \caption{Distribution of median E(V--I) reddening values across the Magellanic system; we choose right ascension for the x-axis to clearly delineate the overall spatial behavior of the system. In the top panel, blue squares are binned E(V--I) values for the HI data from Figure \ref{RRLHI}, converted via Equations \ref{eq10} and \ref{eq11}. Median binned reddenings derived via RR Lyrae from Section \ref{extinction} are shown in cyan. In the bottom panel, the \protect\cite{Schlafly:2011} foreground reddening from the COBE and DIRBE dust maps are shown as gray points and the RR Lyrae are again shown in cyan. In both panels, vertical lines indicates the centers of the LMC (left) and SMC (right). The errorbars represent the standard error of the mean in each bin.}
  \label{RRLreds}
\end{figure}

The reddening from the HI can now be compared to the RRL reddening across the Magellanic system, as seen in the top panel Figure \ref{RRLreds}. The HI content (blue squares) peaks near the centers of the LMC and SMC, and is lower but non-zero in the Bridge region. The reddening inferred from the RR Lyrae suggest greater extinction is present than accounted for by the HI content alone. However, HI does not explicitly trace dust content, although it is often taken as a proxy for the amount of extinction. CO observations of the broader region of the MB would give significant constraints on the dust content between the two galaxies. While presently CO observations are sparse in the Bridge region, focusing primarily on the disks of LMC and SMC (\citealt{Wong:2011}), a study by \cite{Muller:2003} has found CO in the Bridge. New CO observations are likely to be expanded with ALMA; comparing the distribution of younger and older stellar populations to the existence of CO in the Bridge could provide direct evidence of star formation in the region.

We also compare the RR Lyrae reddening from Section \ref{extinction} to the foreground reddening from \cite{Schlafly:2011} in the bottom panel of Figure \ref{RRLreds}. In general, the reddening derived from the RR Lyrae follows the general shape of the foreground reddening; however, the high reddening near the centers of the LMC and SMC from the \cite{Schlafly:2011} foreground reddening is not reflected in the RR Lyrae values. Extinction experienced by the RR Lyrae stars above the foreground levels may be consistent with the existence of dust between the two galaxies. This may be consistent with the reddening histogram in Section \ref{extinction}, which displays an overabundance in the reddening distribution at higher E(V--I).

It is worth checking how well the reddening implied by the RR Lyraes and that inferred for the foreground agree. To do this, we compare the \cite{Guldenschuh:2005} method of using RR Lyrae colors to estimate an average reddening of the Galactic globular clusters to the reddening estimates of \cite{Schlafly:2011} (assuming E(B$-$V) = E(V$-$I)/1.32). The comparisons are provided in Table \ref{tab:redcomps}. While there is scatter, the average offset between the two methods is only 0.009 $\pm$ 0.023 (standard deviation) $\pm$ 0.008 (sem) mag, suggesting that the average RR Lyrae reddening is a reasonable tracer of extinction. This test lends credence to our observation that in the Magellanic system the estimation of foreground reddening from \cite{Schlafly:2011} underestimates the reddening traced by RR Lyrae in LMC. This indicates either the foreground extinction is consistently underestimating the reddening, or that extra dust is present in the MB, leading to additional reddening of RR Lyrae beyond the foreground reddening.

\begin{table*}
\centering
    \begin{minipage}{180mm}
    \caption{Comparison of \protect\cite{Guldenschuh:2005} and \protect\cite{Schlafly:2011} Reddening of Galactic Globular Clusters}
    \begin{tabular}{@{}lcccl@{}}
    \hline
\textbf{Cluster}  & \textbf{E(B$-$V)$_{RR Lyrae}$} & \textbf{E(B$-$V)$_{SF11}$} & \textbf{Difference} & \textbf{Reference} \\  
\hline
NGC 2808 	 &  0.163 	 &  0.1954 	 &  -0.0324		& \cite{Kunder:2013} \\
M 9 		 &  0.425 	 &  0.3746 	 &  0.0504   	& \cite{Arellano-Ferro:2016} \\
NGC 6981 	 &  0.053 	 &  0.0503 	 &  0.0027 		& \cite{Amigo:2013} \\
M 22 	 	&  0.291 	 &  0.2797 	 &  0.0113 		& \cite{Kunder:2013} \\
NGC 3201 	 &  0.23 	 &  0.219 	 &  0.011 		& \cite{Arellano-Ferro:2016} \\
NGC 4590 	 &  0.05 	 &  0.0526 	 &  0.0026 		& \cite{Kains:2015} \\
M 4 	 	&  0.421 	 &  0.4285 	 &  -0.00771 	& \cite{Stetson:2014} \\
M 5 	 	&  0.065 	 &  0.0313 	 &  0.03362	& \cite{Arellano-Ferro:2016} \\
\hline
    \label{tab:redcomps}
    \end{tabular}
    \end{minipage}
\end{table*}

\subsection{Discussion}

If tidal interactions were key in the formation of the MB, we would expect to find an older population of stars in addition to the younger populations. \cite{Bagheri:2013, Skowron:2014, Noel:2015} all present evidence of an older population of stars throughout the Bridge based on CMD analyses. The OGLE observations support this as well, with the presence of RR Lyrae variables in the Bridge confirming the residence of an old (at least 10 Gyr) stellar population. Figures \ref{OGLEcov} and \ref{RRLHI} show that RR Lyrae appear to have a symmetric, non-clumpy distribution between the LMC and SMC, contrary to that of the younger populations.

Additionally, the analysis in Figures \ref{RRLHI} and \ref{RRLreds} do not present evidence that the RR Lyrae and HI intensity follow a coincident spatial distribution. Figure \ref{RRLHI} does not reveal any clear association between areas of greater HI and RR Lyrae positions, supporting the idea that the old population is due in part to an overlapping of the two galactic halos. Figure \ref{RRLreds} shows that the RR Lyrae population is affected by reddening beyond that traced by the foreground reddening. 

Further, our analysis of the properties of the RR Lyrae in the bridge also suggest a gradual mixing of the LMC and SMC populations with distance between the galaxies. We find a clear, smoothly transitioning Magellanic Bridge between the LMC and SMC, as well as a transition from the relatively metal-rich LMC system to the slightly more metal-poor SMC system. These results provide evidence that the Bridge is made up of stars from both the LMC and SMC systems, similar to results from HIPASS data finding the Bridge HI content is likely contributed from both LMC and SMC material (\citealt{Putman:2000}). This is also suggested by \cite{Jacyszyn-Dobrzeniecka:2016}, whose work indicates that RR Lyrae in the Magellanic Bridge region are from the overlapping halo distributions of the LMC and SMC through further exploration of the 3-D structure of the entire Magellanic system.

The sum of these results indicate that the older stellar populations in the MB were likely not affected by the same process as the younger populations, which are associated with HI and clumpy in their distribution across the MB. The older populations could be present in the MB due to either tidal stripping or the extended, overlapping stellar haloes of the LMC and SMC. The reddenings of the RR Lyrae indicate additional dust may be present in the bridge; dust could have been stripped with the old RR Lyrae from their host galaxies. The lack of correlation between HI and RR Lyrae positions implies this is not likely to be the case. Alternatively, the smooth transition in metallicity and distance could result from overlapping halo distributions of the LMC and SMC, which could explain a symmetric, smooth distribution of the old stellar populations.

Large-scale spectroscopic studies will be necessary to determine the LMC or SMC origins of individual stars; this would allow a determination of whether the old populations are overlapping haloes or stripped stars. Additionally, CO observations will be able to more accurately trace areas of high extinction and determine if there is any association with the distribution of stellar populations.

%%%%%%%%%%%%%%%%%%%% Conclusion

\section{Conclusions}\label{conclusions}

The new release of Phase IV from the OGLE survey is used to determine metallicities, reddenings, and distances of RR Lyrae stars in the LMC and SMC system. The analysis examines the metallicity gradients and extinction properties of both galaxies, including examining the properties of the Magellanic bridge and the kinematic properties of the Magellanic system. We present the following conclusions:

1. We derive distances, metallicities, and extinction for ab-type RR Lyrae in the Magellanic Bridge via established methodologies.

2. Examination of the distances between the LMC and SMC show a smooth transition in the distance and metallicities of the old stellar population between the LMC and SMC, connecting the two in a clearly defined Magellanic Bridge.

3. There is little coincidence between the clumpy HI intensity and smooth RR Lyrae positions. This indicates the older population of stars in the bridge may be better explained by the overlap of the LMC and SMC haloes rather than tidally stripped stars.

%%%%%%%%%%%%%%%%%%%%%

\section*{Acknowledgments}

We thank the anonymous referee whose comments and discussion were very helpful. Additionally, we would like to express our appreciation to Mary Putman for providing the HI images from the Parkes telescope.

\bibliographystyle{mn2e}
\bibliography{OGLEIV}

\begin{thebibliography}{68}
\expandafter\ifx\csname natexlab\endcsname\relax\def\natexlab#1{#1}\fi

\bibitem[{{Alcock} {et~al}\mbox{.}(2000){Alcock}, {Allsman}, {Alves},
  {Axelrod}, {Basu}, {Becker}, {Bennett}, {Cook}, {Drake}, {Freeman}, {Geha},
  {Griest}, {King}, {Lehner}, {Marshall}, {Minniti}, {Nelson}, {Peterson},
  {Popowski}, {Pratt}, {Quinn}, {Stubbs}, {Sutherland}, {Tomaney}, {Vandehei},
  \& {Welch}}]{Alcock:2000}
{Alcock} C. {et~al.}, 2000, \aj, 119, 2194

\bibitem[{{Amigo} {et~al}\mbox{.}(2013){Amigo}, {Stetson}, {Catelan},
  {Zoccali}, \& {Smith}}]{Amigo:2013}
{Amigo} P., {Stetson} P.~B., {Catelan} M., {Zoccali} M., {Smith} H.~A., 2013,
  \aj, 146, 130

\bibitem[{{Arellano Ferro} {et~al}\mbox{.}(2016){Arellano Ferro}, {Ahumada},
  {Kains}, \& {Luna}}]{Arellano-Ferro:2016}
{Arellano Ferro} A., {Ahumada} J.~A., {Kains} N., {Luna} A., 2016, \mnras, 461,
  1032

\bibitem[{{Bagheri}, {Cioni} \& {Napiwotzki}(2013){Bagheri}, {Cioni}, \&
  {Napiwotzki}}]{Bagheri:2013}
{Bagheri} G., {Cioni} M.-R.~L., {Napiwotzki} R., 2013, \aap, 551, A78

\bibitem[{{Besla} {et~al}\mbox{.}(2012){Besla}, {Kallivayalil}, {Hernquist},
  {van der Marel}, {Cox}, \& {Kere{\v s}}}]{Besla:2012}
{Besla} G., {Kallivayalil} N., {Hernquist} L., {van der Marel} R.~P., {Cox}
  T.~J., {Kere{\v s}} D., 2012, \mnras, 421, 2109

\bibitem[{{Besla} {et~al}\mbox{.}(2016){Besla}, {Mart{\'{\i}}nez-Delgado}, {van
  der Marel}, {Beletsky}, {Seibert}, {Schlafly}, {Grebel}, \&
  {Neyer}}]{Besla:2016}
{Besla} G., {Mart{\'{\i}}nez-Delgado} D., {van der Marel} R.~P., {Beletsky} Y.,
  {Seibert} M., {Schlafly} E.~F., {Grebel} E.~K., {Neyer} F., 2016, \apj, 825,
  20

\bibitem[{{Bhardwaj} {et~al}\mbox{.}(2014){Bhardwaj}, {Kanbur}, {Singh}, \&
  {Ngeow}}]{Bhardwaj:2014}
{Bhardwaj} A., {Kanbur} S.~M., {Singh} H.~P., {Ngeow} C.-C., 2014, \mnras, 445,
  2655

\bibitem[{{Bohlin}, {Savage} \& {Drake}(1978){Bohlin}, {Savage}, \&
  {Drake}}]{Bohlin:1978}
{Bohlin} R.~C., {Savage} B.~D., {Drake} J.~F., 1978, \apj, 224, 132

\bibitem[{{Borissova} {et~al}\mbox{.}(2006){Borissova}, {Minniti}, {Rejkuba},
  \& {Alves}}]{Borissova:2006}
{Borissova} J., {Minniti} D., {Rejkuba} M., {Alves} D., 2006, \aap, 460, 459

\bibitem[{{Borissova} {et~al}\mbox{.}(2004){Borissova}, {Minniti}, {Rejkuba},
  {Alves}, {Cook}, \& {Freeman}}]{Borissova:2004}
{Borissova} J., {Minniti} D., {Rejkuba} M., {Alves} D., {Cook} K.~H., {Freeman}
  K.~C., 2004, \aap, 423, 97

\bibitem[{{Brocato} {et~al}\mbox{.}(1996){Brocato}, {Castellani}, {Ferraro},
  {Piersimoni}, \& {Testa}}]{Brocato:1996}
{Brocato} E., {Castellani} V., {Ferraro} F.~R., {Piersimoni} A.~M., {Testa} V.,
  1996, \mnras, 282, 614

\bibitem[{{Chaboyer}(1999)}]{Chaboyer:1999}
{Chaboyer} B., 1999, Post-Hipparcos Cosmic Candles, 237, 111

\bibitem[{{Chen} {et~al}\mbox{.}(2014){Chen}, {Indebetouw}, {Muller},
  {Kawamura}, {Gordon}, {Sewi{\l}o}, {Whitney}, {Fukui}, {Madden}, {Meade},
  {Meixner}, {Oliveira}, {Robitaille}, {Seale}, {Shiao}, \& {van
  Loon}}]{Chen:2014}
{Chen} C.-H.~R. {et~al.}, 2014, \apj, 785, 162

\bibitem[{{Collinge}, {Sumi} \& {Fabrycky}(2006){Collinge}, {Sumi}, \&
  {Fabrycky}}]{Collinge:2006}
{Collinge} M.~J., {Sumi} T., {Fabrycky} D., 2006, \apj, 651, 197

\bibitem[{{de Grijs} \& {Bono}(2015)}]{de-Grijs:2015}
{de Grijs} R., {Bono} G., 2015, \aj, 149, 179

\bibitem[{{de Grijs}, {Wicker} \& {Bono}(2014){de Grijs}, {Wicker}, \&
  {Bono}}]{de-Grijs:2014}
{de Grijs} R., {Wicker} J.~E., {Bono} G., 2014, \aj, 147, 122

\bibitem[{{Deb} \& {Singh}(2014)}]{Deb:2014}
{Deb} S., {Singh} H.~P., 2014, \mnras, 438, 2440

\bibitem[{{Diaz} \& {Bekki}(2012)}]{Diaz:2012}
{Diaz} J.~D., {Bekki} K., 2012, \apj, 750, 36

\bibitem[{{Diplas} \& {Savage}(1994)}]{Diplas:1994}
{Diplas} A., {Savage} B.~D., 1994, \apj, 427, 274

\bibitem[{{Dorfi} \& {Feuchtinger}(1999)}]{Dorfi:1999}
{Dorfi} E.~A., {Feuchtinger} M.~U., 1999, \aap, 348, 815

\bibitem[{{Fukui} {et~al}\mbox{.}(2014){Fukui}, {Okamoto}, {Kaji}, {Yamamoto},
  {Torii}, {Hayakawa}, {Tachihara}, {Dickey}, {Okuda}, {Ohama}, {Kuroda}, \&
  {Kuwahara}}]{Fukui:2014}
{Fukui} Y. {et~al.}, 2014, \apj, 796, 59

\bibitem[{{Gonidakis} {et~al}\mbox{.}(2009){Gonidakis}, {Livanou}, {Kontizas},
  {Klein}, {Kontizas}, {Belcheva}, {Tsalmantza}, \&
  {Karampelas}}]{Gonidakis:2009}
{Gonidakis} I., {Livanou} E., {Kontizas} E., {Klein} U., {Kontizas} M.,
  {Belcheva} M., {Tsalmantza} P., {Karampelas} A., 2009, \aap, 496, 375

\bibitem[{{Grondin}, {Demers} \& {Kunkel}(1992){Grondin}, {Demers}, \&
  {Kunkel}}]{Grondin:1992}
{Grondin} L., {Demers} S., {Kunkel} W.~E., 1992, \aj, 103, 1234

\bibitem[{{Guldenschuh} {et~al}\mbox{.}(2005){Guldenschuh}, {Layden}, {Wan},
  {Whiting}, {van der Bliek}, {Baca}, {Carlin}, {Freismuth}, {Mora}, {Salyk},
  {Vera}, {Verdugo}, \& {Young}}]{Guldenschuh:2005}
{Guldenschuh} K.~A. {et~al.}, 2005, \pasp, 117, 721

\bibitem[{{Haschke}, {Grebel} \& {Duffau}(2011){Haschke}, {Grebel}, \&
  {Duffau}}]{Haschke:2011}
{Haschke} R., {Grebel} E.~K., {Duffau} S., 2011, \aj, 141, 158

\bibitem[{{Haschke}, {Grebel} \& {Duffau}(2012){Haschke}, {Grebel}, \&
  {Duffau}}]{Haschke:2012}
{Haschke} R., {Grebel} E.~K., {Duffau} S., 2012, \aj, 144, 106

\bibitem[{{Haschke} {et~al}\mbox{.}(2012){Haschke}, {Grebel}, {Duffau}, \&
  {Jin}}]{Haschke:2012a}
{Haschke} R., {Grebel} E.~K., {Duffau} S., {Jin} S., 2012, \aj, 143, 48

\bibitem[{{Inno} {et~al}\mbox{.}(2016){Inno}, {Bono}, {Matsunaga},
  {Fiorentino}, {Marconi}, {Lemasle}, {da Silva}, {Soszy{\'n}ski}, {Udalski},
  {Romaniello}, \& {Rix}}]{Inno:2016}
{Inno} L. {et~al.}, 2016, ArXiv e-prints

\bibitem[{{Jacyszyn-Dobrzeniecka} {et~al}\mbox{.}(2016){Jacyszyn-Dobrzeniecka},
  {Skowron}, {Mr{\'o}z}, {Soszy{\'n}ski}, {Udalski}, {Pietrukowicz}, {Skowron},
  {Poleski}, {Koz{\l}owski}, {Wyrzykowski}, {Pawlak}, {Szyma{\'n}ski}, \&
  {Ulaczyk}}]{Jacyszyn-Dobrzeniecka:2016}
{Jacyszyn-Dobrzeniecka} A.~M. {et~al.}, 2016, ArXiv e-prints

\bibitem[{{Jeffery} {et~al}\mbox{.}(2011){Jeffery}, {Smith}, {Brown},
  {Sweigart}, {Kalirai}, {Ferguson}, {Guhathakurta}, {Renzini}, \&
  {Rich}}]{Jeffery:2011}
{Jeffery} E.~J. {et~al.}, 2011, \aj, 141, 171

\bibitem[{{Johnson} {et~al}\mbox{.}(1999){Johnson}, {Bolte}, {Stetson},
  {Hesser}, \& {Somerville}}]{Johnson:1999}
{Johnson} J.~A., {Bolte} M., {Stetson} P.~B., {Hesser} J.~E., {Somerville}
  R.~S., 1999, \apj, 527, 199

\bibitem[{{Jurcsik}(1995)}]{Jurcsik:1995}
{Jurcsik} J., 1995, \actaa, 45, 653

\bibitem[{{Kains} {et~al}\mbox{.}(2015){Kains}, {Arellano Ferro}, {Figuera
  Jaimes}, {Bramich}, {Skottfelt}, {J{\o}rgensen}, {Tsapras}, {Street},
  {Browne}, {Dominik}, {Horne}, {Hundertmark}, {Ipatov}, {Snodgrass}, {Steele},
  {Lcogt/Robonet Consortium}, {Alsubai}, {Bozza}, {Calchi Novati}, {Ciceri},
  {D'Ago}, {Galianni}, {Gu}, {Harps{\o}e}, {Hinse}, {Juncher}, {Korhonen},
  {Mancini}, {Popovas}, {Rabus}, {Rahvar}, {Southworth}, {Surdej}, {Vilela},
  {Wang}, {Wertz}, \& {Mindstep Consortium}}]{Kains:2015}
{Kains} N. {et~al.}, 2015, \aap, 578, A128

\bibitem[{{Kallivayalil} {et~al}\mbox{.}(2013){Kallivayalil}, {van der Marel},
  {Besla}, {Anderson}, \& {Alcock}}]{Kallivayalil:2013}
{Kallivayalil} N., {van der Marel} R.~P., {Besla} G., {Anderson} J., {Alcock}
  C., 2013, \apj, 764, 161

\bibitem[{{Kapakos} \& {Hatzidimitriou}(2012)}]{Kapakos:2012}
{Kapakos} E., {Hatzidimitriou} D., 2012, \mnras, 426, 2063

\bibitem[{{Kinman} {et~al}\mbox{.}(1991){Kinman}, {Stryker}, {Hesser},
  {Graham}, {Walker}, {Hazen}, \& {Nemec}}]{Kinman:1991}
{Kinman} T.~D., {Stryker} L.~L., {Hesser} J.~E., {Graham} J.~A., {Walker}
  A.~R., {Hazen} M.~L., {Nemec} J.~M., 1991, \pasp, 103, 1279

\bibitem[{{Kunder}, {Chaboyer} \& {Layden}(2010){Kunder}, {Chaboyer}, \&
  {Layden}}]{Kunder:2010}
{Kunder} A., {Chaboyer} B., {Layden} A., 2010, \aj, 139, 415

\bibitem[{{Kunder} {et~al}\mbox{.}(2013){Kunder}, {Stetson}, {Cassisi},
  {Layden}, {Bono}, {Catelan}, {Walker}, {Paredes Alvarez}, {Clem},
  {Matsunaga}, {Salaris}, {Lee}, \& {Chaboyer}}]{Kunder:2013}
{Kunder} A. {et~al.}, 2013, \aj, 146, 119

\bibitem[{{Layden}, {Anderson} \& {Husband}(2013){Layden}, {Anderson}, \&
  {Husband}}]{Layden:2013}
{Layden} A., {Anderson} T., {Husband} P., 2013, ArXiv e-prints

\bibitem[{{Layden} \& {Sarajedini}(2000)}]{Layden:2000}
{Layden} A.~C., {Sarajedini} A., 2000, \aj, 119, 1760

\bibitem[{{Mizuno} {et~al}\mbox{.}(2006){Mizuno}, {Muller}, {Maeda},
  {Kawamura}, {Minamidani}, {Onishi}, {Mizuno}, \& {Fukui}}]{Mizuno:2006}
{Mizuno} N., {Muller} E., {Maeda} H., {Kawamura} A., {Minamidani} T., {Onishi}
  T., {Mizuno} A., {Fukui} Y., 2006, \apjl, 643, L107

\bibitem[{{Muller}, {Staveley-Smith} \& {Zealey}(2003){Muller},
  {Staveley-Smith}, \& {Zealey}}]{Muller:2003}
{Muller} E., {Staveley-Smith} L., {Zealey} W.~J., 2003, \mnras, 338, 609

\bibitem[{{No{\"e}l} {et~al}\mbox{.}(2013){No{\"e}l}, {Conn}, {Carrera},
  {Read}, {Rix}, \& {Dolphin}}]{Noel:2013}
{No{\"e}l} N.~E.~D., {Conn} B.~C., {Carrera} R., {Read} J.~I., {Rix} H.-W.,
  {Dolphin} A., 2013, \apj, 768, 109

\bibitem[{{No{\"e}l} {et~al}\mbox{.}(2015){No{\"e}l}, {Conn}, {Read},
  {Carrera}, {Dolphin}, \& {Rix}}]{Noel:2015}
{No{\"e}l} N.~E.~D., {Conn} B.~C., {Read} J.~I., {Carrera} R., {Dolphin} A.,
  {Rix} H.-W., 2015, \mnras, 452, 4222

\bibitem[{{Olsen} {et~al}\mbox{.}(1998){Olsen}, {Hodge}, {Mateo}, {Olszewski},
  {Schommer}, {Suntzeff}, \& {Walker}}]{Olsen:1998}
{Olsen} K.~A.~G., {Hodge} P.~W., {Mateo} M., {Olszewski} E.~W., {Schommer}
  R.~A., {Suntzeff} N.~B., {Walker} A.~R., 1998, \mnras, 300, 665

\bibitem[{{Papadakis} {et~al}\mbox{.}(2000){Papadakis}, {Hatzidimitriou},
  {Croke}, \& {Papamastorakis}}]{Papadakis:2000}
{Papadakis} I., {Hatzidimitriou} D., {Croke} B.~F.~W., {Papamastorakis} I.,
  2000, \aj, 119, 851

\bibitem[{{Pejcha} \& {Stanek}(2009)}]{Pejcha:2009}
{Pejcha} O., {Stanek} K.~Z., 2009, \apj, 704, 1730

\bibitem[{{Putman}(2000)}]{Putman:2000}
{Putman} M.~E., 2000, \pasa, 17, 1

\bibitem[{{Putman} {et~al}\mbox{.}(2003){Putman}, {Staveley-Smith}, {Freeman},
  {Gibson}, \& {Barnes}}]{Putman:2003}
{Putman} M.~E., {Staveley-Smith} L., {Freeman} K.~C., {Gibson} B.~K., {Barnes}
  D.~G., 2003, \apj, 586, 170

\bibitem[{{R\r{u}{\v z}i{\v c}ka}, {Theis} \& {Palou{\v s}}(2009){R\r{u}{\v
  z}i{\v c}ka}, {Theis}, \& {Palou{\v s}}}]{Ruuzicka:2009}
{R\r{u}{\v z}i{\v c}ka} A., {Theis} C., {Palou{\v s}} J., 2009, \apj, 691, 1807

\bibitem[{{Saha} {et~al}\mbox{.}(2010){Saha}, {Olszewski}, {Brondel}, {Olsen},
  {Knezek}, {Harris}, {Smith}, {Subramaniam}, {Claver}, {Rest}, {Seitzer},
  {Cook}, {Minniti}, \& {Suntzeff}}]{Saha:2010}
{Saha} A. {et~al.}, 2010, \aj, 140, 1719

\bibitem[{{Sarajedini}(2011)}]{Sarajedini:2011}
{Sarajedini} A., 2011, in RR Lyrae Stars, Metal-Poor Stars, and the Galaxy,
  {McWilliam} A., ed., Vol.~5, p. 181

\bibitem[{{Schlafly} \& {Finkbeiner}(2011)}]{Schlafly:2011}
{Schlafly} E.~F., {Finkbeiner} D.~P., 2011, \apj, 737, 103

\bibitem[{{Skowron} {et~al}\mbox{.}(2014){Skowron}, {Jacyszyn}, {Udalski},
  {Szyma{\'n}ski}, {Skowron}, {Poleski}, {Koz{\l}owski}, {Kubiak},
  {Pietrzy{\'n}ski}, {Soszy{\'n}ski}, {Mr{\'o}z}, {Pietrukowicz}, {Ulaczyk}, \&
  {Wyrzykowski}}]{Skowron:2014}
{Skowron} D.~M. {et~al.}, 2014, \apj, 795, 108

\bibitem[{{Skowron} {et~al}\mbox{.}(2016){Skowron}, {Soszy{\'n}ski}, {Udalski},
  {Szyma{\'n}ski}, {Pietrukowicz}, {Poleski}, {Wyrzykowski}, {Ulaczyk},
  {Koz{\l}owski}, {Skowron}, {Mr{\'o}z}, \& {Pawlak}}]{Skowron:2016}
{Skowron} D.~M. {et~al.}, 2016, ArXiv e-prints

\bibitem[{{Smith}(1995)}]{Smith:1995}
{Smith} H.~A., 1995, Cambridge Astrophysics Series, 27

\bibitem[{{Smolec}(2005)}]{Smolec:2005}
{Smolec} R., 2005, \actaa, 55, 59

\bibitem[{{Soszy{\'n}ski} {et~al}\mbox{.}(2009){Soszy{\'n}ski}, {Udalski},
  {Szyma{\'n}ski}, {Kubiak}, {Pietrzy{\'n}ski}, {Wyrzykowski}, {Szewczyk},
  {Ulaczyk}, \& {Poleski}}]{Soszynski:2009}
{Soszy{\'n}ski} I. {et~al.}, 2009, \actaa, 59, 1

\bibitem[{{Soszy{\'n}ski} {et~al}\mbox{.}(2016){Soszy{\'n}ski}, {Udalski},
  {Szyma{\'n}ski}, {Wyrzykowski}, {Ulaczyk}, {Poleski}, {Pietrukowicz},
  {Koz{\l}owski}, {Skowron}, {Skowron}, {Mr{\'o}z}, \&
  {Pawlak}}]{Soszynski:2016}
{Soszy{\'n}ski} I. {et~al.}, 2016, ArXiv e-prints

\bibitem[{{Stetson} {et~al}\mbox{.}(2014){Stetson}, {Braga}, {Dall'Ora},
  {Bono}, {Buonanno}, {Ferraro}, {Iannicola}, {Marengo}, \&
  {Neeley}}]{Stetson:2014}
{Stetson} P.~B. {et~al.}, 2014, \pasp, 126, 521

\bibitem[{{Udalski} {et~al}\mbox{.}(2012){Udalski}, {Kowalczyk},
  {Soszy{\'n}ski}, {Poleski}, {Szyma{\'n}ski}, {Kubiak}, {Pietrzy{\'n}ski},
  {Koz{\l}owski}, {Pietrukowicz}, {Ulaczyk}, {Skowron}, \&
  {Wyrzykowski}}]{Udalski:2012}
{Udalski} A. {et~al.}, 2012, \actaa, 62, 133

\bibitem[{{Udalski} {et~al}\mbox{.}(2003){Udalski}, {Pietrzynski}, {Szymanski},
  {Kubiak}, {Zebrun}, {Soszynski}, {Szewczyk}, \& {Wyrzykowski}}]{Udalski:2003}
{Udalski} A., {Pietrzynski} G., {Szymanski} M., {Kubiak} M., {Zebrun} K.,
  {Soszynski} I., {Szewczyk} O., {Wyrzykowski} L., 2003, \actaa, 53, 133

\bibitem[{{Udalski} {et~al}\mbox{.}(2008){Udalski}, {Szymanski}, {Soszynski},
  \& {Poleski}}]{Udalski:2008}
{Udalski} A., {Szymanski} M.~K., {Soszynski} I., {Poleski} R., 2008, \actaa,
  58, 69

\bibitem[{{Udalski}, {Szyma{\'n}ski} \& {Szyma{\'n}ski}(2015){Udalski},
  {Szyma{\'n}ski}, \& {Szyma{\'n}ski}}]{Udalski:2015}
{Udalski} A., {Szyma{\'n}ski} M.~K., {Szyma{\'n}ski} G., 2015, \actaa, 65, 1

\bibitem[{{van der Marel} \& {Cioni}(2001)}]{vanderMarel:2001}
{van der Marel} R.~P., {Cioni} M.-R.~L., 2001, \aj, 122, 1807

\bibitem[{{Wagner-Kaiser} \& {Sarajedini}(2013)}]{Wagner-Kaiser:2013}
{Wagner-Kaiser} R., {Sarajedini} A., 2013, \mnras, 431, 1565

\bibitem[{{Wong} {et~al}\mbox{.}(2011){Wong}, {Hughes}, {Ott}, {Muller},
  {Pineda}, {Bernard}, {Chu}, {Fukui}, {Gruendl}, {Henkel}, {Kawamura},
  {Klein}, {Looney}, {Maddison}, {Mizuno}, {Paradis}, {Seale}, \&
  {Welty}}]{Wong:2011}
{Wong} T. {et~al.}, 2011, \apjs, 197, 16

\bibitem[{{Zinn} \& {West}(1984)}]{Zinn:1984}
{Zinn} R., {West} M.~J., 1984, \apjs, 55, 45

\end{thebibliography}
\clearpage

\label{lastpage}

\end{document}